\documentclass[twocolumn, secnumarabic,amssymb, nobibnotes, aps, prd]{revtex4-2}
\usepackage{hyperref}
\usepackage{amsmath}
\usepackage{graphics}
\usepackage{graphicx}
\usepackage{float}  
\usepackage{caption}

\begin{document}
	\title{Dark Matter \& Dark Energy in the Three-Higgs Doublet Model}%
	
	\author{Mohid Farhan}%
	\email[Mohid Farhan: ]{mohidf35@gmail.com}
	\affiliation{Department of Space Science, Institute of Space Technology, Islamabad, Pakistan}
	
		\author{Ibtehaj Hassan}%
	\email[Ibtehaj Hassan: ]{ibtehaj.hassan@ist.edu.pk}
	\affiliation{Department of Space Science, Institute of Space Technology, Islamabad, Pakistan}
	
		\author{Muhammad Usman}%
	\email[Muhammad Usman: ]{usman.muhammad@ist.edu.pk}
	\affiliation{Department of Space Science, Institute of Space Technology, Islamabad, Pakistan}
	
	\author{Noraiz Tahir}%
	\email[Noraiz Tahir:]{noraiz.tahir@sns.nust.edu.pk}
	\affiliation{Department of Physics and Astronomy, School of Natural Sciences (SNS), National University of Sciences and Technology (NUST), Islamabad, Pakistan}
	
	\date{\today}%

	\begin{abstract}
		This article discusses the incorporation of dark matter and dark energy into a new physics model called the Three-Higgs Doublet Model. Dark matter and dark energy are accommodated as CP-even and $Z_2$-odd scalars in their respective inert doublets. By leveraging a $Z_2$ symmetry to suppress certain interactions, we model the behavior of dark matter. Similarly, by imposing a shift symmetry, dark energy can be mimicked within the same framework for the current cosmic epoch. The dark matter relic density is calculated for our model using \texttt{micrOMEGAs}. It is shown that despite the inclusion of dark energy, dark matter relic density can be brought within observational bounds and match existing literature. Furthermore, the one-loop and two-loop Renormalization Group Equations (RGEs) were computed using \texttt{SARAH} to ensure radiative stability over a large range of energies. This study lays the groundwork for a future study of dark matter-dark energy interactions in the early universe and the exploration of different early universe dynamics.
	\end{abstract}
	\maketitle
	\tableofcontents
	
	\section{Introduction}
	As it stands, dark matter and dark energy are often introduced as discrepancy-filling entities, reflecting gaps in our current understanding of the universe at the fundamental level. Among the most profound of these discrepancies are Galactic Rotation Curves (GRCs) \cite{Persic1996} and the accelerated expansion of the universe \cite{Riess1998}. The ideas of separate sectors of the universe consisting of dark matter and dark energy attempt to explain the former and latter, respectively. Observations indicate that baryonic matter only comprises about 5\% of the energy contents of the universe with Cold Dark Matter (CDM) making up approximately 25 \% and dark energy attributing to approximately 70\% of the universe's matter-energy contents. It is worth mentioning that there also exist alternate explanations for the aforementioned anomalies, the most popular of which are Modification of Newtonian Dynamics (MOND) \cite{Sanders2002} and f($\mathbf{R}$) gravity \cite{Sotiriou2010}. These theories have been prone to deficiencies in explaining other phenomena like the excessive gravitational lensing around galaxy clusters \cite{Clowe2006}. There also exist many other alternate theories for gravity as well but we will assume the existence of CDM and dark energy.

	In cosmology, dark energy appears in the form of cosmological constant $\Lambda$ which was initially introduced to invoke “negative” pressure in an attempt to counter gravitational collapse, and retain a static universe. Having been discarded following Hubble's discovery of an expanding universe, the cosmological constant reappeared when it was deduced that this expansion was accelerating \cite{Riess1998}. There are a a number of scalar field models, the most prominent being the quintessence, in which the equation of state remains in between $\frac{1}{3}$ and -1. In this article, we take dark energy as a quintessence field whose zero-point energy is
	$
	\rho_{\Lambda} \simeq 10^{-47}~\text{GeV}^4
	$
	which leads to a mass scale of	$m \simeq 10^{-33}~$\text{eV}. The generic quintessence potential is 	\begin{equation}
		V(\phi) = \frac{1}{2} m^2 \phi^2+\lambda \phi^4,
	\end{equation}
	where \(\phi\) represents a field displacement of the Planck scale~\cite{Copeland2006}. Due to this ultra-light scalar mass, most particle physics models suffer from fine-tuning problems, which we address here. 
	
	The standard model (SM) of particle physics gained strong experimental support with the discovery of the Higgs boson in 2012 at the Large Hadron Collider by the A Toroidal LHC ApparatuS (ATLAS) and Compact Muon Solenoid (CMS) collaborations~\cite{atlas2012,cms2012}. Despite this success, several anomalies persist, among them is the problem of dark matter and dark energy. These challenges indicate that the SM is incomplete, motivating the development of new theories that extend it. One of the simplest extensions is the Inert Doublet Model (IDM) \cite{LopezHonorez2007} where a $Z_2$ symmetry is invoked. This symmetry distinguishes between contents of the active Higgs doublet and the proposed inert doublet and allows the lightest interacting particle to be stable for long periods of time due to its non coupling with fermions. Under this symmetry, these lightest interacting particle become Weakly Interacting Massive Particles (WIMPs) as $Z_2$ forbids these particles from decaying into W and Z boson. Here, we assume the existence of WIMPs as candidates for dark matter. 
	
	Grzadkowski et. al. in 2009 proposed the extension of IDM where an additional active scalar doublet was introduced to enable Charge-Parity (CP) violation~\cite{Grzadkowski2009}. Note that when the CP operator is applied to a system of particles, the coordinates of the particles are mirrored, and the particles are converted to anti-particles. The violation of this symmetry allows for the explanation of matter-antimatter asymmetry. This variant is referred to as the I(1+2)IDM. Furthermore, Usman et. al. in 2018 showed that dark energy can be modeled into the IDM \cite{UsmanQadir2018}, motivating us to seek a dark energy candidate in the extended models of standard model. Motivated by the existence of three fermion families, an extension known as I(2+1)IDM was introduced in a series of works from 2014 to 2018, primarily led by Keus et. al.~\cite{Keus2015}. This model includes an additional inert scalar doublet, allowing CP violation in the non-inert sector and improving the testability of the dark matter candidate in the mass ranges $53~\text{GeV} \lesssim m \lesssim m_W$ (mass of the W boson, which is $\approx$80.3 GeV) and $m_{\text{DM}} \gtrsim 525~\text{GeV}$, which is relevant to our work. The I(2+1)IDM provides two inert doublets in addition to the standard model Higgs doublet. These inert doublets provide a natural home for the dark matter and dark energy scalars. In the present epoch, the dark sector is constrained by a strict global $Z_2$ symmetry and an additional shift symmetry acting on the dark energy scalar ($H_1^0$). $Z_2$
	symmetry ensures the stability of dark matter by forbidding couplings to fermions, while the shift symmetry forbids polynomial interactions of the dark energy scalar with other scalars ensuring its inertness. In the strict $g_1,g_2 \rightarrow 0$ limit, the dark energy doublet behaves effectively as a gauge singlet. We retain the doublet structure to accommodate the possibility of nonzero gauge interactions in the early universe, while in the present epoch the phenomenology reduces to that of a singlet. 
	
	The article is as follows in the next section we discuss the theoretical framework where the particle content is defined along with the scalar potential. The vacuum expectation values are shown after finding the minima of the potential. 
	The Yukawa interaction terms are displayed and mass matrices are diagonalized. Then, the dark matter relic density is computed using \texttt{micrOMEGAs}. After briefly explaining the computation procedure and defining parameters like $\delta$ and $\Delta$, the plots are shown in the section \ref{sec-obs} 
	and a viable mass region is obtained that is consistent with \textit{Planck's} data \cite{planck2018}. Later in sec. (\ref{sec-rad}) to ensure radiative stability due to the ultralight scalar of dark energy, the constraints on its quartic couplings are derived using the Coleman-Weinberg potential. The higher-loop renormalizable group equations are computed using \texttt{SARAH}. An epoch-dependent gauge limit ($g_1,g_2 \rightarrow 0$) will also be introduced to address the fine-tuning concerns. Section (\ref{sec-conclusion}) is the conclusion and discussion.
	
	\section{Theoretical Framework}
	The 3HDM allows for rich phenomenology. It is centered on two inert doublets, $H_1$ and $H_2$. These doublets are unable to participate in Electroweak Symmetry Breaking (EWSB) since they do not acquire a vacuum expectation value. The active standard model Higgs doublet is denoted by $H_3$. The doublets are mathematically defined as:
	\begin{equation}
		H_1 = \begin{pmatrix}
			H_1^{+} \\
			\frac{H_1^0 + i A_1^0}{\sqrt{2}}
		\end{pmatrix}, H_2 = \begin{pmatrix}
			H_2^{+} \\
			\frac{H_2^0 + i A_2^0}{\sqrt{2}}
		\end{pmatrix}, H_3 = \begin{pmatrix}
			G^{+} \\
			\frac{v + h + i G^0}{\sqrt{2}}
		\end{pmatrix}.
	\end{equation}
	Here, $H_3$ denotes the standard model-like Higgs doublet with a vacuum expectation value $v$, and accommodates the observed 125~GeV Higgs boson $h$ ~\cite{atlas2012,cms2012}. The associated Goldstone bosons, $G^0$ and $G^+$, are absorbed as the longitudinal components of the $Z$ and $W^+$ gauge bosons, respectively. Conversely, $H_1$
	and $H_2$ do not acquire vacuum expectation values. This renders them stable (or long-lived) and forms the inert sector, which naturally accommodates dark matter candidates \cite{LopezHonorez2007}. $H_1^+$ and $H_2^+$ are charged pseudo-scalars. The neutral components $H_1^0$ and $H_2^0$ are CP-even candidates, and $A_1^0$ and $A_2^0$ are CP-odd scalars.

	The $Z_2$ symmetry helps distinguish between the active and inert doublets as they transform as:
	$$H_1 \rightarrow -H_1$$
	$$H_2 \rightarrow -H_2$$
	$$H_3 \rightarrow H_3$$
	
    The invariance of the Lagrangian under the sign inversion of $H_1$ and $H_2$ physically implies the inert nature of these doublets. This $Z_2$ symmetry also stabilizes the dark matter candidate, which in our model is the CP-even and $Z_2$-odd scalar ($H_2^0$). The dark energy (quintessence) scalar field ($H_1^0$) mimics the cosmological constant in the current cosmic epoch, as stated earlier.
	
	The scalar potential of the I(2+1)IDM is given by:
\[
V_H = -\mu_1^2 \left(H_1^\dagger H_1 \right) -\mu_2^2 \left(H_2^\dagger H_2\right)  -\mu_3^2 \left(H_3^\dagger H_3 \right)
\]
\[
+\lambda_{11}\left(H_1^\dagger H_1\right)^2 +\lambda_{22}\left(H_2^\dagger H_2\right)^2 +\lambda_{33}\left(H_3^\dagger H_3\right)^2
\]
\[
+\lambda_{12}\left(H_1^\dagger H_1\right)\left(H_2^\dagger H_2\right)+\lambda_{23}\left(H_2^\dagger H_2\right)\left(H_3^\dagger H_3\right)
\]
\[
+\lambda_{31}\left(H_3^\dagger H_3\right)\left(H_1^\dagger H_1\right) + \lambda^{'}_{12}\left(H_1^\dagger H_2\right)\left(H_2^\dagger H_1\right)
\]
\[
+\lambda^{'}_{23}\left(H_2^\dagger H_3\right)\left(H_3^\dagger H_2\right)+\lambda^{'}_{31}\left(H_3^\dagger H_1\right)\left(H_1^\dagger H_3\right) 
\]
\[ 
	-\mu^2_{12}\left(H_1^\dagger H_2 \right)+\lambda_1 \left( H_1^\dagger H_2 \right)^2 +\lambda_2 \left( H_2^\dagger H_3 \right)^2 
\]
\begin{equation}
 + \lambda_3 \left( H_3^\dagger H_1 \right)^2 + \text{h.c.}
\end{equation}
The scalar potential $V_H$ includes all renormalizable, gauge-invariant terms involving three Higgs doublets $H_1$, $H_2$, and $H_3$. The quadratic terms $\mu_i^2$ ($i = 1,2,3$) control the mass scales associated with the respective doublets. The parameters $\lambda_{ii}$ represent self-interactions of each doublet, while the $\lambda_{ij}$ ($i \neq j$) terms describe quartic interactions between the norms of different doublets. The $\lambda'_{ij}$ terms are Hermitian products and encode additional interactions that preserve the norm but not the phase symmetry between the fields. The soft-breaking mass term $\mu_{12}^2$ breaks any potential discrete symmetry between $H_1$ and $H_2$, while the terms $\lambda_1$, $\lambda_2$, and $\lambda_3$ introduce complex phases via $(H_i^\dagger H_j)^2$-type interactions and may be sources of CP violation. The Hermitian conjugates of these complex terms (denoted by h.c.) ensure the potential remains Hermitian.

The minimum of this potential can be calculated by solving the following set of equations.
\[v_1 v_2^2 Re\left( \lambda_1 \right) +v_1 v_3^2 Re\left( \lambda_3 \right) + v_1^3 \lambda_{11} + \frac{1}{2} v_1 v_2^2 \lambda_{12}+ \frac{1}{2} v_1 v_3^2 \lambda_{31} \]
\begin{equation}
	- v_1 \mu_1^2 - v_2 Re\left( \mu_{12}^2 \right) + \frac{1}{2} v_1 v_2^2 \lambda_{12}^{'}+ \frac{1}{2} v_1 v_3^2 \lambda_{13}^{'}=0
\end{equation}
\[v_1^2 v_2 Re\left( \lambda_1\right) + v_2 v_3^2 Re\left( \lambda_2\right) + \frac{1}{2} v_1^2 v_2 \lambda_{12} + v_2^3 \lambda_{22} + \frac{1}{2} v_2 v_3^2 \lambda_{23} \]
\begin{equation}
	- v_2 \mu_2^2 - v_1 Re\left( \mu_{12}^2\right) + \frac{1}{2} v_1^2 v_2 \lambda_{12}^{'} + \frac{1}{2} v_2 v_3^2 \lambda_{23}^{'}=0
\end{equation}
\[v_2^2 v_3 \lambda_2 + v_1^2 v_3 \lambda_3 + \frac{1}{2} v_2^2 v_3 \lambda_{23} + \frac{1}{2} v_1^2 v_3 \lambda_{31}\]
\begin{equation}
	+ v_3^3 \lambda_{33} - v_3 \mu_3^2 + \frac{1}{2} v_2^2 v_3 \lambda_{23}^{'} + \frac{1}{2} v_1^2 v_3 \lambda_{31}^{'}=0
\end{equation}
Here we choose the solution, 
\begin{equation}
	v_1 = 0,v_2 = 0,v_3 = \sqrt{\frac{\mu_3^2}{\lambda_{33}}}.
\end{equation}
which makes perfect sense since $H_1$ and $H_2$, due to inertness, do not acquire a vacuum expectation value. Since $H_3$ is the only active doublet, Yukawa Lagrangian in this model is identical to the SM Yukawa Lagrangian as \cite{Yukawa:1935xg}:
 \begin{equation}
 	\mathcal{L}_{\text{Yukawa}} = 
 	- \left[ 
 	y_d\, \bar{Q}_L \Phi\, d_R 
 	+ y_u\, \bar{Q}_L \tilde{\Phi}\, u_R 
 	+ y_e\, \bar{L}_L \Phi\, e_R 
 	+ \text{h.c.}
 	\right]
 \end{equation}
The mass matrices are given by,
\begin{equation}
	M_{H^{\pm}}^2 = \begin{pmatrix}
		\mu_1^2 - \frac{\lambda_{31} \mu_3^2}{2\lambda_{33}} & \mu_{12}^2 \\
		\mu_{12}^2 & \mu_2^2 - \frac{\lambda_{23} \mu_3^2}{2\lambda_{33}}
	\end{pmatrix}
\end{equation}
\begin{align}
	M_H^2 &= 
	\resizebox{0.36\textwidth}{!}{$
		\begin{pmatrix}
			\mu_1^2 
			- \frac{\lambda_3 \mu_3^2}{\lambda_{33}} 
			- \frac{(\lambda_{31} + \lambda^{'}_{31}) \mu_3^2}{2\lambda_{33}} 
			& \mu_{12}^2 \\
			\mu_{12}^2 
			& \mu_2^2 
			- \frac{\lambda_2 \mu_3^2}{\lambda_{33}} 
			- \frac{(\lambda_{23} + \lambda^{'}_{23}) \mu_3^2}{2\lambda_{33}}
		\end{pmatrix}
		$} \label{eq:MH2} \\[6pt]
	M_A^2 &= 
	\resizebox{0.36\textwidth}{!}{$
		\begin{pmatrix}
			\mu_1^2 
			+ \frac{\lambda_3 \mu_3^2}{\lambda_{33}} 
			- \frac{(\lambda_{31} + \lambda^{'}_{31}) \mu_3^2}{2\lambda_{33}} 
			& \mu_{12}^2 \\
			\mu_{12}^2 
			& \mu_2^2 
			+ \frac{\lambda_2 \mu_3^2}{\lambda_{33}} 
			- \frac{(\lambda_{23} + \lambda^{'}_{23}) \mu_3^2}{2\lambda_{33}}
		\end{pmatrix}
		$} \label{eq:MA2}
\end{align}

and $m_h^2 = 2\mu_3^2$ , $m_{G^0}^2=0$ and $m_{G^{\pm}}^2=0$. The mass matrices can be diagonalized by using,
\begin{equation}
	H_1^0 = Cos\theta_h {H}_1 - Sin\theta_h {H}_2,
\end{equation}
\begin{equation}
	H_2^0 = Sin\theta_h {H}_1 - Cos\theta_h {H}_2,
\end{equation}
\[
tan2\theta_h = \frac{2\mu_{12}^2}{\left( \mu_1^2 - \frac{\lambda_3 \mu_3^2}{\lambda_{33}} - \frac{\lambda_{31} \mu_3^2}{2\lambda_{33}}  - \frac{\mu_3^2 \lambda_{31}^{'}}{2\lambda_{33}}\right)} 
\]
\begin{equation}
	\frac{ }{- \left( \mu_2^2 - \frac{\lambda_2 \mu_3^2}{\lambda_{33}} - \frac{\lambda_{23} \mu_3^2}{2\lambda_{33}} - \frac{\mu_3^2 \lambda_{23}^{'}}{2\lambda_{33}}\right)},
\end{equation}
\begin{equation}
	H_1^{\pm} = Cos\theta_c {H}_1^{\pm} - Sin\theta_c {H}_2^{\pm},
\end{equation}
\begin{equation}
	H_2^{\pm} = Sin\theta_c {H}_1^{\pm} - Cos\theta_c {H}_2^{\pm},
\end{equation}
\begin{equation}
	tan2\theta_c = \frac{2\mu_{12}^2}{\left( \mu_1^2 - \frac{\lambda_{31} \mu_3^2}{2\lambda_{33}}\right)- \left( \mu_2^2 - \frac{\lambda_{23} \mu_3^2}{2\lambda_{33}}\right)}, 
\end{equation}
\begin{equation}
	A_1^{0} = Cos\theta_a {A}_1 - Sin\theta_a {A}_2,
\end{equation}
\begin{equation}
	A_2^{0} = Sin\theta_a {A}_1 - Cos\theta_a {A}_2,
\end{equation}
\[
tan2\theta_a = \frac{2\mu_{12}^2}{\left( \mu_1^2 + \frac{\lambda_3 \mu_3^2}{\lambda_{33}} - \frac{\lambda_{31} \mu_3^2}{2\lambda_{33}}  - \frac{\mu_3^2 \lambda_{31}^{'}}{2\lambda_{33}}\right)} 
\]
\begin{equation}
	\frac{ }{- \left( \mu_2^2 + \frac{\lambda_2 \mu_3^2}{\lambda_{33}} - \frac{\lambda_{23} \mu_3^2}{2\lambda_{33}} - \frac{\mu_3^2 \lambda_{23}^{'}}{2\lambda_{33}}\right)}.
\end{equation}

Mass of inert particles are given by,
\[m_{H_1}^2=-Cos^2 \theta_h  \mu_1^2 - Sin^2 \theta_h  \mu_2^2 + \frac{Sin^2 \theta_h  \lambda_2 \mu_3^2}{\lambda_{33}}\]
\[ + \frac{Cos^2 \theta_h \lambda_3 \mu_3^2}{\lambda_{33}} + \frac{Sin^2 \theta_h \lambda_{23} \mu_3^2}{2\lambda_{33}} + \frac{Cos^2 \theta_h \lambda_{31} \mu_3^2}{2\lambda_{33}}\]
\begin{equation}
	- 2Cos\theta_h Sin\theta_h \mu_{12}^2 + \frac{Sin^2 \theta_h \mu_3^2 \lambda_{23}^{'}}{2\lambda_{33}} + \frac{Cos^2 \theta_h \mu_3^2 \lambda_{31}^{'}}{2\lambda_{33}},
\end{equation}

\[m_{H_2}^2 = - Sin^2\theta \mu_1^2 - Cos^2\theta \mu_2^2 + \frac{Cos^2\theta \lambda_2 \mu_3^2}{\lambda_{33}} + \frac{Sin^2\theta \lambda_3 \mu_3^2}{\lambda_{33}}\]

\begin{equation}
	+ \frac{Cos^2\theta \lambda_{23} \mu_3^2}{2\lambda_{33}} + \frac{Sin^2\theta \lambda_{31} \mu_3^2}{2\lambda_{33}} +\frac{Cos^2\theta \mu_3^2 \lambda_{23}^{'}}{2\lambda_{33}} + \frac{Sin^2\theta \mu_3^2 \lambda_{31}^{'}}{2\lambda_{33}}.
\end{equation}

It is clear that if $H_1^0$ is a candidate for the Quintessence field then self-couplings and all the scalar and Gauge couplings of $H_1$ must be absent. This is not suitable for dark matter in $53 GeV \leq m_{DM} \leq m_W$ mass region because the authors of \cite{Keus:2014jha} have used the $\textbf{H}_2 \textbf{A}_2 \rightarrow Z$ and $\textbf{H}_1 \textbf{H}_1 \rightarrow h$ co-annihilation channels to fulfill Planck, LUX and LHC constraints on dark matter mass, given that $\Delta = m_{H_2} - m_{H_1}$ is small. In the heavy mass region $m_{DM} \geq 525 GeV$, $\Delta \geq 50 GeV$ leads to viable DM relic density values. Exact values of $\Delta$ above $~ 50 GeV$ does not make any significant difference in the relic density calculations. In this case, $H_1$ and $H_2$ are effectively decoupled and $\textbf{H}_1,\textbf{A}_1,\textbf{H}_1^{\pm}$ do not affect the dark matter relic density calculations. This decoupling reduces I(2+1)IDM to IDM. This is presented as “Case H” in \cite{Keus2015}, where $\Delta$ is kept large and mass of the dark matter candidate is varied.

\section{Observational Constraints on dark matter relic density}\label{sec-obs}
The main dark matter relic density constraint imposed on our model is based on the observed values of dark matter relic density which has an acceptable range as follows \cite{planck2018}:
\begin{equation}
	\Omega h^2=0.119 \pm 0.0027,
\end{equation}
with $1\sigma$ accuracy of results, where $\Omega$ is the density parameter and h is the reduced Hubble's constant.
We take a similar approach as \cite{Keus2015}, and aim to obtain a parameter space which respects \cite{planck2018} for viability. The coupling constant between the Higgs boson and the dark matter candidate, $\lambda_L$, is also constrained to a low value.

We used \texttt{micrOMEGAs}(Version-6.1.15) to carry out the dark matter relic density calculations \cite{Belanger2010}. The tool numerically solves the Boltzmann equation, which is a first-order differential equation that computes the relic density:
\begin{equation}
	\frac{dn_s}{dt} = - 3Hn_s -\langle \sigma v \rangle ( n_s^2 - n_{s,eq}^2 ),
\end{equation}
where  $n_s$ is the number density, H is the Hubble parameter which accounts for the expansion of the universe. $\langle \sigma v \rangle$ is the annihilation cross section where $\sigma$ is the cross-section and v is the average velocity. $(n_s^2 - n_{s,eq}^2)$ depicts how far the system is from thermal equilibrium. $n_s$ is related with mass density $\rho$ by:
\begin{equation}
	\rho = m_{DM} n_s,
\end{equation}
The density parameter of the universe $\Omega$ is given by:
\begin{equation}
	\Omega=\frac{\rho}{\rho_c},
\end{equation}   
where $\rho_c$ is the critical density of the universe which is the density required to have a geometrically flat universe:
\begin{equation}
	\rho_c=\frac{3H^2}{8\pi G},
\end{equation}
The product of the density parameter $\Omega$ and the reduced Hubble constant h is the relic density. Once $n_s$ is computed, the dark matter mass remains the only free variable involved in the computation of dark matter relic density. Among the relevant parameters are:
$$\Delta=m_{H_2^0}-m_{H_1^0},$$
$$\delta=m_{H_2^0}-m_{A_2^0},$$
The mass of the dark energy scalar is negligible compared to the mass of dark matter, reducing the I(2+1)IDM to the IDM due to a large $\Delta$. The annihilation cross section is also heavily dependent on $\delta$ which is the mass splitting between CP-even and CP-odd scalar of $H_2$. 

\begin{figure}[H]
	\centering
	\includegraphics[width=1\linewidth]{"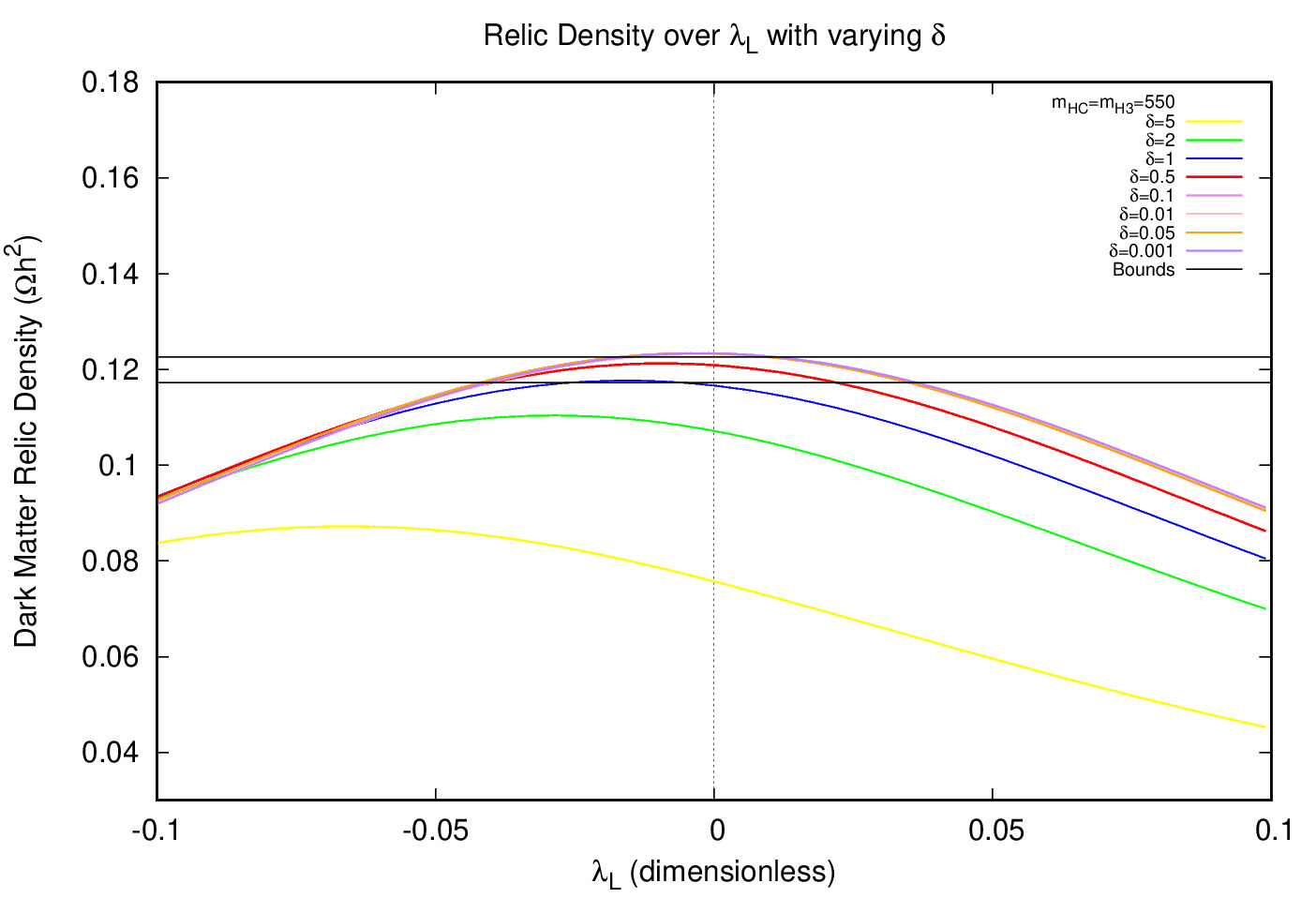"}
	\caption{Dark matter relic density as a function of the coupling constant for different values of $\delta$.}
	\label{fig:fig}
\end{figure}

This dependency is demonstrated in FIG. 1 which shows the variance of relic density when masses are kept constant and $\delta$ is varied. For low $\delta$, the lower the coupling constant between dark matter and the Higgs boson, and vice versa. $\Omega h^2$ decreases as the coupling constant increases due to the reactive nature of relic density before freeze-out, resulting in more dark matter particles decaying into standard model particles. Note that $\delta$ cannot be less than zero as it violates our assumption of dark matter candidate being a CP-even scalar. In such a case, the CP-even scalar becomes lighter than the CP-odd scalar and could decay into lighter components, hence, losing its stability. A key observation is that for $m_{DM} \approx550$ GeV, the relic density begins to fall within observational bounds as $\delta \leq 1$. In this regime, the relic density curves are close to each other, as is shown in FIG. 2.

\begin{figure}[H]
	\centering
	\includegraphics[width=1\linewidth]{"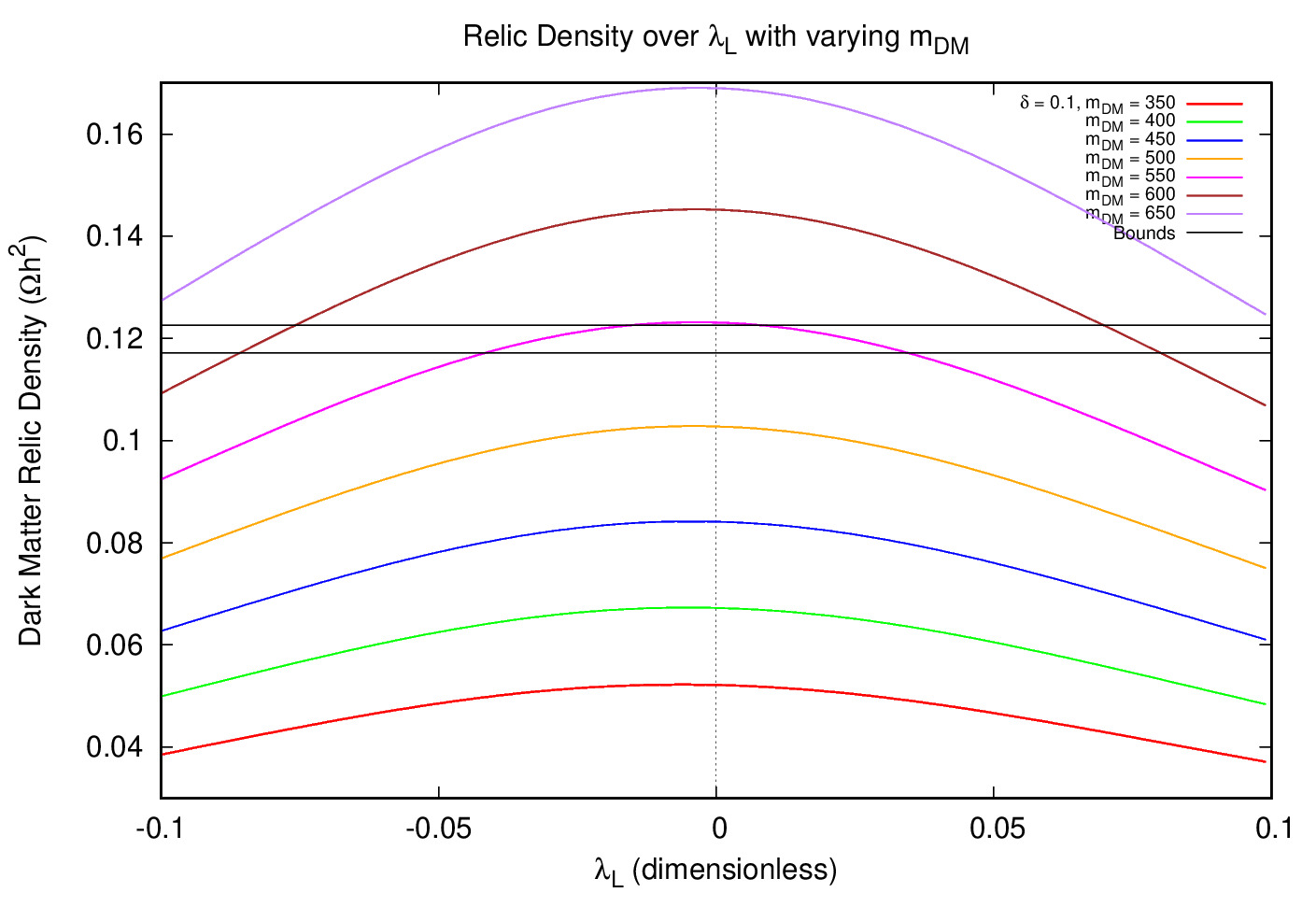"}
	\caption{Dark matter relic density plotted as a function of $\lambda_L$ for varying dark matter mass ($m_{DM}$).}
	\label{fig:fig}
\end{figure}

FIG. 2 shows $\Omega h^2$ plotted against $\lambda_L$ for a large $\Delta$ value. This vast difference between the doublets, effectively decouples the $H_1$. We set $\delta=0.1$ and the mass of dark matter was varied. FIG. 1 justifies the $\delta$ value for FIG. 2, as it shows that all $\delta$ values below $\delta=0.1$ output, effectively, the same curves and so there is no impact on dark matter relic density in the region $0<\delta<0.1$.

Here, we notice a similar trend as FIG. 1, where relic density peaks when coupling is at its weakest and drops as the coupling constant increases to either side, since more dark matter particles decay into SM particles. It also shows that higher dark matter mass corresponds to a higher relic density. This is to be expected if $n_s$ is approximately constant (see eq. 25). We can now vary the mass of dark matter so that the relic density fits within bounds imposed by eq. (23). Using FIG. 2, a region corresponding to the lower bound of $m_{DM}$ can be found at approximately $m_{DM} \approx 550$ GeV by varying $m_{DM}$ with increments for 5 GeV along $520 \leq m_{DM} \leq 550$. To determine the mass of dark matter for which the observational bounds are respected, we used trial and error to determine the exact mass (up until a whole value of GeV) which is shown by the two red curves in FIG. 2. 

For $m_{DM} \leq 535$, the highest value of dark matter relic density still falls short of the constraint in eq. (23), and for $m_{DM} \geq 549$, the relic density is too high to be within the confines of eq. (23). Note that theoretically, any value satisfying $m_{DM} \geq 536$ GeV can yield relic density that obeys eq. (23). However, as $m_{DM} \geq 549$ GeV , the window of the permissible coupling constant drastically reduces. For heavier dark matter candidates, the coupling constant also needs to increase to compensate for the extra mass that increases the overall relic density. So, the heavier the candidate, the more reactive it needs to be with the SM particles to reach the required relic density constraints. Since the couplings constant has been experimentally constrained to a value near zero \cite{atlas2012, cms2012}, this is not a viable parameter space.

However, there is a narrow region of dark matter mass that respects constraint  in eq. (23) without causing discontinuity in the magnitude of the coupling constant. This is a very viable window since introducing discontinuity not only becomes observationally problematic, but also introduces fine-tuning problems as observations would only be fulfilled for very specfic values of $m_{DM}$. This narrow region shown by FIG. 2, is in between the	$536 \leq m_{DM} \leq 548$ GeV . 

The authors of \cite{Keus2015} use the coupling constant $g_{H1H1h}$ on the x-axis of their analysis. To directly compare our results with theirs, we will rescale the x-axis using the relation $g_{H_1 H_1 h} = 2\lambda_L v$ as given in  \cite{LopezHonorez2007}. After rescaling, we scanned across $m_{DM}$ for $\delta=0.5$ GeV (FIG. 3) and $\delta=1$ GeV (FIG. 4). 
\begin{figure}[H]
	\centering
	\includegraphics[width=1\linewidth]{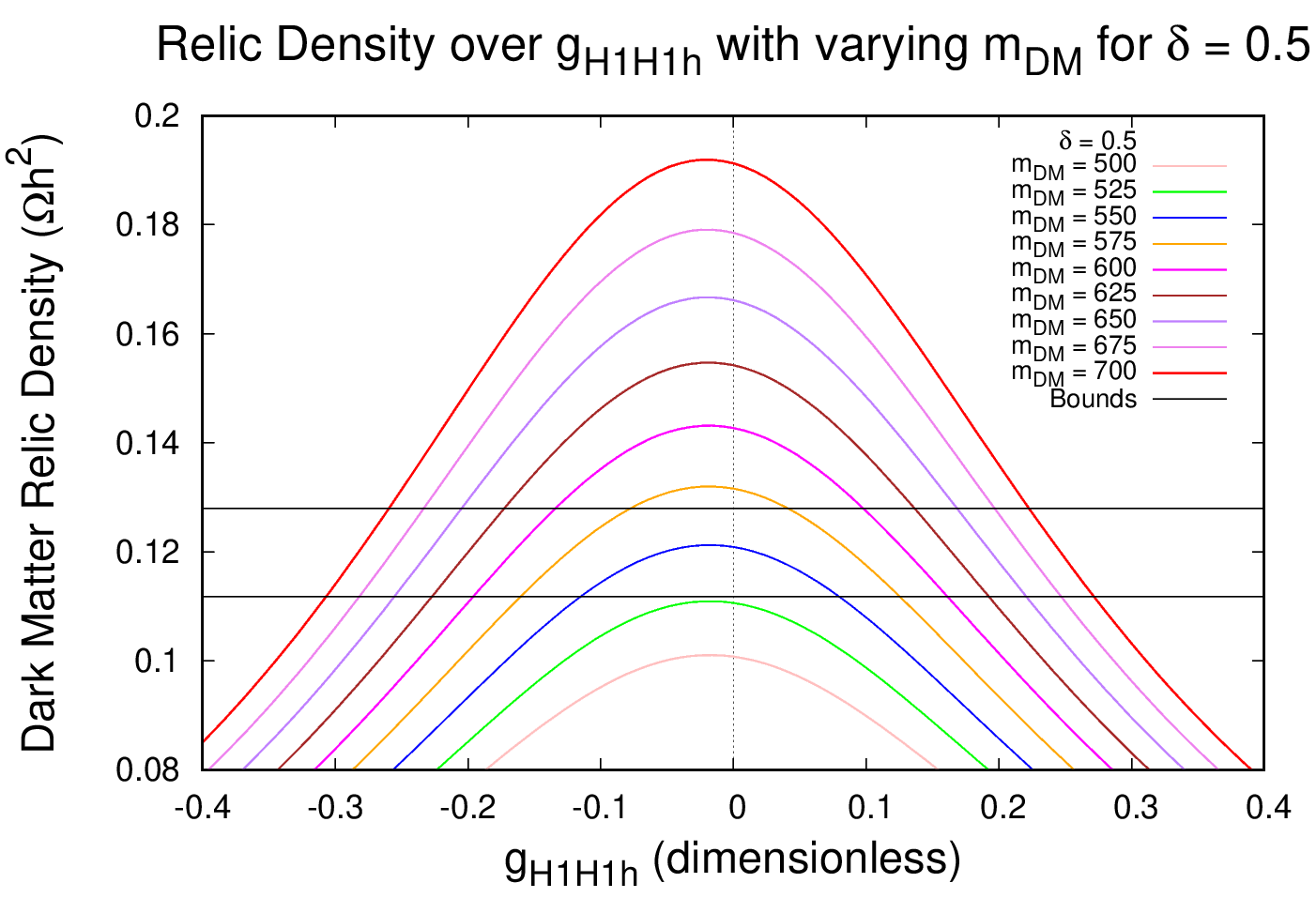}
	\caption{This plot varies $m_{DM}$ from 500 GeV to 700 GeV with increments of 25 GeV, while setting $\delta=0.5$. It is a direct comparison to figure 6 (right) presented in \cite{Keus2015}.}
\end{figure}

\begin{figure}[H]
	\centering[H]
	\includegraphics[width=1\linewidth]{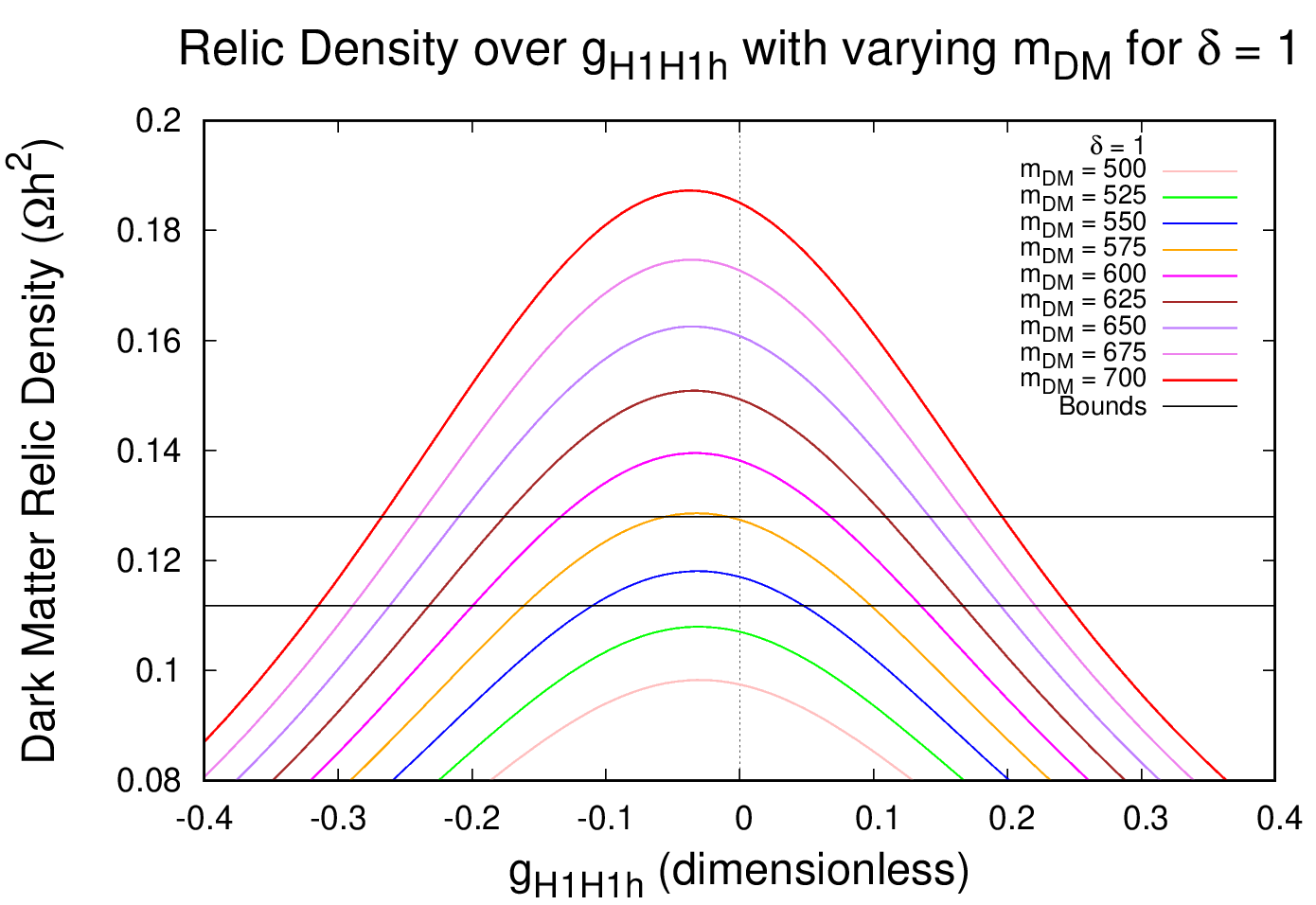}
	\caption{This plot varies $m_{DM}$ from 500 GeV to 700 GeV with increments of 25 GeV, while setting $\delta=1$. It is a direct comparison to case H presented in \cite{Keus2015}. }

\end{figure}

In both FIG. 3 and FIG. 4, $\Delta$ was kept high. From FIG. 4, we conclude that our findings agree with the established literature in \cite{Keus2015}. This serves as a green-flag for the viability of our model. Note that for full consistency, we increased the bound from eq. (23) to 3$\sigma$. This was made possible due to the fact that $\Delta$ values above 50 GeV do not affect the dark matter relic density as supported by \cite{Keus2015}. The reason for this is the decoupling of $H_1$, which is a direct consequence of Boltzmann suppression:
$$\frac{n_{H_1}}{n_{H_2}} \approx e^{\frac{m_{H_1}-m_{H_2}}{T}}\approx e^{\frac{-\Delta}{T}},$$
As the mass deficit $\Delta$ increases, this ratio goes to zero and so does the thermally averaged annihilation cross-section ($\langle \sigma_{eff} v \rangle$) since:
$$\langle \sigma_{eff} v \rangle = \sum_{ij} \langle \sigma_{ij} v_{ij} \rangle 
\frac{n_i^\text{eq} n_j^\text{eq}}{n_S^\text{eq} n^\text{eq}_S}$$
Therefore, after a threshold value of $\Delta$ (in this case 50 GeV), the relic density curves come out to be unaffected due to the negligible contribution of ${H_1}$. With the findings for FIG. 4, our model sits firmly on established literature.

\section{Radiative Stability of the Dark Energy scalar}\label{sec-rad}

The stability of the 3HDM hinges on robustness against radiative effects. Most of the general stability constraints have been laid out in \cite{Keus:2014ipa}. The uniqueness of our model comes from the addition of dark energy in three-Higgs Doublet Model. 
Due to the ultralight dark energy scalar, we must also account for radiative corrections, otherwise known as quantum loop corrections. Generally, the effects of radiative corrections are suppressed by tiny couplings. This supports model's validity for the current cosmic epoch. However, the ultralight scalar of dark energy introduces some constraints. For scalars, one can compute the one-loop potential using the Coleman-Weinberg potential \cite{Coleman:1973jx}. The one-loop correction to the scalar potential is given by:

\begin{equation}
	V^{(1)}(\phi) = \frac{1}{64\pi^2} \sum_i (-1)^{2s_i}(2s_i + 1) m_i^4(\phi) \left( \log\frac{m_i^2(\phi)}{\mu^2} - \frac{3}{2} \right),
\end{equation}

where \( s_i \) is the spin of particle \( i \), \( m_i^2(\phi) \) is the field-dependent mass of particle \( i \), and \( \mu \) is the renormalization scale. Since we are dealing with a scalar field, spin $s$ can be set to zero, simplifying the expression. $m_i \approx 10^{-42}$ GeV as shown in \cite{Copeland2006}. The first-order corrected mass is given by:

\begin{equation}  
	m_i^2(\phi)=m_i^2+ \alpha s^2,
\end{equation} 
where $\alpha$ is the effective quartic coupling. To compute the  mass correction due to first order loop corrections, we utilize the condition:

\begin{equation}  
	\delta m_{H}^2 = \left. \frac{\partial^2 V^{(1)}(s)}{\partial s^2} \right|_{s=0}.
\end{equation} 

The full form of the Coleman-Weinberg potential for the dark energy scalar is given by:

	\begin{multline}
		V^{(1)}(s) = \frac{1}{64\pi^{2}} \Big[
		(m_{h}^{2} + \alpha_{h}s^{2})^{2} \left( \log\left( \frac{m_{h}^{2} + \alpha_{h}s^{2}}{\mu^{2}} \right) - \frac{3}{2} \right) \\
		+ (m_{H_{1}}^{2} + \alpha_{h1}s^{2})^{2} \left( \log\left( \frac{m_{H_{1}}^{2} + \alpha_{h1}s^{2}}{\mu^{2}} \right) - \frac{3}{2} \right) \\
		+ (m_{H_{2}}^{2} + \alpha_{h2}s^{2})^{2} \left( \log\left( \frac{m_{H_{2}}^{2} + \alpha_{h2}s^{2}}{\mu^{2}} \right) - \frac{3}{2} \right)
		\Big],
	\end{multline}

where $m_h$ is mass of Higgs boson, $\alpha_h$ is the effective quartic coupling to the Higgs portal, and $\alpha_{h1}$ and $\alpha_{h2}$ are the effective self quartic coupling and the quartic coupling to dark matter, respectively.

Applying eq.(31) to eq.(30), we get (See Appendix A):
\begin{equation}
	\delta m_{H_1}^2 = \frac{1}{16\pi^2} \, \alpha_i m_{h}^2 \left( \log\left( \frac{m_{h}^2}{\mu^2} \right) - 1 \right),
\end{equation}
This reduction is possible by neglecting dark energy self-couplings and dark energy-dark matter couplings, which is valid for the current cosmic epoch. Hence, the only term relevant to us is the heavier Higgs field that couples with dark energy. From the scalar potential (eq. 3), we see that $\alpha_{h}$ can be written as:
$$\alpha_{h}
\approx \lambda_{31} \approx \lambda_{31}^{\prime},$$ 
Furthermore, $\mu$ is a free parameter which is set to a value comparable to the Higgs mass. This choice of $\mu$ matches the energy scale of electroweak symmetry breaking, at which the corrections to the dark energy scalar would be most prominent, hence $\mu=m_{h}$. Therefore, our expression simplifies to: 
\[
\delta m_{H_1}^2 \approx - \frac{\lambda_{31} m_h^2}{16\pi^2},
\]
Note that we assume $\lambda_{31}$ to evolve similarly to $\lambda_{31}^{\prime}$ and to avoid redundancy, we will proceed to use only 
$\lambda_{31}$. To nullify significant quantum corrections, the ultralight dark energy scalar mass constraint demands that:

$$|\delta m_{H_1}^2| \leq 10^{-84} GeV^2$$ 

Putting in \( m_h = 125 \) GeV, and rearranging for $\lambda_{31}$ we obtain the order-of-magnitude constraint:
\begin{equation}
	\lambda_{31} \leq 10^{-86}
\end{equation}
For our dark energy scalar to be radiatively stable, this condition must be fulfilled, which raises fine-tuning concerns. However, 't Hooft's criterion \cite{tHooft:1979bh} states that a tiny value for a parameter is technically natural if setting it to zero restores a symmetry. In our case, this is the shift symmetry imposed solely on the dark energy scalar $H_1^0$ \cite{tHooft:1979bh}:
$$H_1^0 \rightarrow H_1^0+c,$$

The extension of this symmetry to the full doublet is unnecessary, since only $H_1^0$ is relevant for cosmological dynamics.  Similar shift-symmetry approaches have been widely applied in quintessence and pNGB DE models \cite{vasilev2024, low2014} lending precedent to our construction. From eq. (3), the terms involving $\lambda_{31}$, $\lambda_{31}^{\prime}$, $\lambda_{12}$, $\lambda_{12}^{\prime}$, $\mu_1^2$, $\mu_{12}^2$, $\lambda_1$ and $\lambda_3$ are all variant under this symmetry. Thereby, setting all these parameters to zero significant increases the symmetry of the theory. As a consequence, the seemingly fine-tuned constraint in eq. (33) is technically natural under the 't Hooft's criterion of technical naturalness \cite{tHooft:1979bh}. Since this shift symmetry can be applied if the gauge couplings are negligible, the local gauge symmetry is not broken and in the $g_1,g_2\rightarrow0$, the DE doublet acts as a singlet in the current epoch. This should be understood as an effective description in the present epoch; full gauge invariance of the 3HDM is preserved at high scales where $g_1,g_2$ are nonzero. We have retained the 3HDM structure to allow for DM-DE interactions in the early universe, which are currently negligible due to the imposed shift symmetry.

To further investigate the stability of the dark energy scalar and the naturalness of a small $\lambda_{31}$, we compute the RGEs for this parameter. The renormalizable group equations were calculated for $\lambda_{31}$ using a \texttt{mathematica} package named \texttt{SARAH}. We allowed for the violation of the global $Z_2$ symmetry for demonstration. The one-loop $\beta$ function was computed to be:
\begin{equation}
	\begin{aligned}
		\beta^{(1)}_{\lambda_{31}} =\ 
		& \frac{27}{100} g_1^4 - \frac{9}{10} g_1^2 g_2^2 + \frac{9}{4} g_2^4 
		- \frac{9}{5} g_1^2 \lambda_{31} - 9 g_2^2 \lambda_{31} \\
		& + 12 \lambda_{11} \lambda_{31} + 4 \lambda_{11} \lambda_{31}' 
		+ 4 \lambda_{12} \lambda_{23} + 2 \lambda_{12} \lambda_{23}' \\
		&+ 2 \lambda_{12}' \lambda_{23} 
		 + 4 \lambda_{31}^2 + 12 \lambda_{31} \lambda_{33} 
		+ 2 {\lambda_{31}'}^2 \\
		&+ 4 \lambda_{31}' \lambda_{33} 
		+ 8 \lambda_3  \overline{\lambda_3} 
		 + 6 \lambda_{31} \operatorname{Tr}[Y_d Y_d^\dagger] \\
		& + 2 \lambda_{31} \operatorname{Tr}[Y_e Y_e^\dagger]
		 + 6 \lambda_{31} \operatorname{Tr}[Y_u Y_u^\dagger]
	\end{aligned}
\end{equation}

Here, $g_1$ and $g_2$ denote the gauge couplings associated with the $U(1)_Y$ and $SU(2)_L$ symmetries, respectively. The parameters $\lambda_{ij}$ and $\lambda_{ij}'$ represent quartic couplings between the Higgs doublets $H_i$ and $H_j$, with primed terms denoting distinct cross-coupling structures allowed by the scalar potential. The complex scalar coupling $\lambda_3$ encodes interactions of the form $(H_i^\dagger H_j)^2$, and $\overline{\lambda_3}$ is its complex conjugate. The terms involving traces over Yukawa matrices, such as $\operatorname{Tr}[Y_d Y_d^\dagger]$, arise from fermionic loop corrections, capturing the effects of down-type quarks ($Y_d$), up-type quarks ($Y_u$), and charged leptons ($Y_e$) in the renormalization group running. These flavor-universal contributions reflect the radiative impact of matter fields on the evolution of $\lambda_{31}$ since mathematically:

\begin{equation}
	\mu \frac{d \lambda_{31}}{d\mu} 
	= \beta^{(1)}_{\lambda_{31}} 
	+ \frac{1}{(16\pi^2)} \beta^{(2)}_{\lambda_{31}} 
	+ \frac{1}{(16\pi^2)^2} \beta^{(3)}_{\lambda_{31}} + \cdots
\end{equation}
 where $\mu$ is the energy scale and the superscripts on the $\beta$ function show the order of corrections.

Similar to the one-loop beta function, the two-loop $\beta$ function for $\lambda_{31}$ was also computed and is provided in the supplementary material. Note that under a strict shift symmetry, $\lambda_{31}$ vanishes and so in that case, the beta function also vanishes.

These findings show that in the scenario of a broken or violated $Z_2$ and shift symmetry, gauge terms independent from $\lambda_{31}$ arise in the $\beta$ functions. Therefore, even if $\lambda_{31}$ were set to zero, the radiative corrections would push $\lambda_{31}$ well past its constraint. The computation of these $\beta$ functions was carried out for a scalar potential that included cross-terms with non-zero coupling constants. If one is to repeat this procedure for a strictly conserved shift symmetry, cross-terms involving $H_1$ should not be defined in the scalar potential, thereby implying that couplings $\lambda_{31}$ are automatically set to zero. This gives intuitive sense of the 't Hooft's criterion.

In the absence of shift symmetry and the violation of the $g_1,g_2\rightarrow0$ limit, quantum corrections induced by gauge interactions can regenerate small portal couplings like $\lambda_{31}$ even if they vanish at a given scale. This reflects the additive gauge contributions in the 1-loop and 2-loop beta functions. Therefore, while 't Hooft’s criterion justifies the naturalness of tiny  values in the exact symmetric limit, the stability of such couplings under renormalizable group evolution requires strict enforcement of all shift symmetry-preserving conditions in the Lagrangian. This makes the tiny dark energy coupling technically natural under the 't Hooft's criterion. In this sense, $\lambda_{13}\rightarrow0$ is not merely a fine-tuned choice but corresponds to a symmetry-protected fixed point of the RGEs, stabilized by the imposed shift symmetry.

\section{Conclusion}\label{sec-conclusion}
In this paper, the 3HDM was used to incorporate dark energy into a framework that previously only included dark matter and baryonic matter. The scalar potential for this framework was defined and mass matrices were diagonalized. We showed that the dark matter relic density constraints are fulfilled for a continuous window of $\lambda_L$ in the mass region $536\leq m_{DM} \leq 548$ GeV, given that $\delta=0.1$ by using \textbf{micrOMEGAs}. Furthermore, we demonstrated that due to a large $\Delta$ for the current cosmic epoch, the results agree with already established literature, specifically case H and figure 6 of \cite{Keus2015}.
Finally, the dark energy's radiative stability was analyzed. We derived the constraints on its effective quartic coupling (with the Higgs field) using the Coleman-Weinberg potential and found $\lambda_{31} \leq 10^{-86}$, which introduced fine-tuning concerns. This calculation assumed negligible self interaction and interactions with dark matter in the current cosmic epoch, due to the imposition of the shift symmetry on the dark energy scalar in the negligible gauge coupling limit.

By computing the RGEs using \textbf{SARAH}, we found that the gauge coupling terms in the $\beta$ function of $\lambda_{31}$ push it well above the imposed constraint, if the gauge couplings are too large and the gauge couplings are restored. This points towards a reactive behavior of dark energy in the early universe, which is explorable as future work. To address this concern, we used the $g_1,g_2 \rightarrow 0$ limit for the current cosmic epoch, in which the dark energy doublet reduces phenomenologically to a gauge singlet. This can be interpreted as an effective decoupling of the dark energy scalar from electroweak gauge bosons in the current cosmic epoch. We retain the doublet structure to allow for the possibility of nonzero gauge interactions in the early universe, which is a future direction for research. 

It is important to note that this work was limited to the study of the current cosmic epoch and that this paper lays the groundwork for a future study that caters for the early universe's dynamics. This can be made possible by softly breaking $Z_2$ to activate dark matter interactions, and lifting of the $g_1, g_2 \rightarrow 0$ limit for the dark energy doublet at earlier scales and analyzing the possibility of interactions in the early universe. The one-loop and two-loop beta function offer accurate predictions of loop corrected contributions to the evolution of all relevant parameters involved as they show the evolution of parameters over different energy scales. They can be particularly useful for sensitive parameters and can assist in future work.

\appendix
\section*{Appendix A: Derivation of Mass Correction from the One-Loop Effective Potential}
To differentiate the given one-loop Coleman-Weinberg potential $V^{(1)}(s)$ twice with respect to $s$, we perform the differentiation term by term.

The potential is given by:
\[
V^{(1)}(s)=\frac{1}{64\pi^{2}}\left\{(m_{h}^{2}+\alpha_{h}s^{2})^{2}\left[\log\left(\frac{m_{h}^{2}+\alpha_{h}s^{2}}{\mu^{2}}\right)-\frac{3}{2}\right] \right.
\]
\[
\left. +(m_{H_{1}}^{2}+\alpha_{h1}s^{2})^{2}\left[\log\left(\frac{m_{H_{1}}^{2}+\alpha_{h1}s^{2}}{\mu^{2}}\right)-\frac{3}{2}\right] \right.
\]
\[
\left. +(m_{H_{2}}^{2}+\alpha_{h2}s^{2})^{2}\left[\log\left(\frac{m_{H_{2}}^{2}+\alpha_{h2}s^{2}}{\mu^{2}}\right)-\frac{3}{2}\right]\right\}
\]

Let's consider a generic term in the sum, $f_i(s)$, corresponding to a particle $i$ (e.g., $h$, $H_1$, $H_2$):
\[
f_i(s) = (m_i^2 + \alpha_i s^2)^2 \left[\log\left(\frac{m_i^2 + \alpha_i s^2}{\mu^2}\right) - \frac{3}{2}\right]
\]
To simplify the differentiation, let $X_i(s) = m_i^2 + \alpha_i s^2$. Then $f_i(s) = X_i(s)^2 \left[\log\left(\frac{X_i(s)}{\mu^2}\right) - \frac{3}{2}\right]$.
We also note that $\frac{dX_i}{ds} = 2\alpha_i s$.

We use the chain rule: $\frac{df_i}{ds} = \frac{df_i}{dX_i} \frac{dX_i}{ds}$.

First, differentiate $f_i(X_i)$ with respect to $X_i$:
$$
	\frac{d}{dX_i} \left( X_i^2 \log\left(\frac{X_i}{\mu^2}\right) - \frac{3}{2}X_i^2 \right) $$
	$$= \left( 2X_i \log\left(\frac{X_i}{\mu^2}\right) + X_i^2 \cdot \frac{1}{X_i} \right) - 3X_i $$ 
	$$= 2X_i \log\left(\frac{X_i}{\mu^2}\right) + X_i - 3X_i $$
	$$= 2X_i \left( \log\left(\frac{X_i}{\mu^2}\right) - 1 \right)$$

Now, combine with $\frac{dX_i}{ds}$:
\begin{align*}
	\frac{df_i}{ds} &= \left[ 2X_i \left( \log\left(\frac{X_i}{\mu^2}\right) - 1 \right) \right] \cdot (2\alpha_i s) \\
	&= 4\alpha_i s (m_i^2 + \alpha_i s^2) \left(\log\left(\frac{m_i^2 + \alpha_i s^2}{\mu^2}\right) - 1\right)
\end{align*}

To obtain the second derivative,
we differentiate $\frac{df_i}{ds}$ again using the product rule. Let $A = 4\alpha_i s$, $B = (m_i^2 + \alpha_i s^2)$, and $C = \left(\log\left(\frac{m_i^2 + \alpha_i s^2}{\mu^2}\right) - 1\right)$.
So, $\frac{d^2f_i}{ds^2} = \frac{dA}{ds}BC + A\frac{dB}{ds}C + AB\frac{dC}{ds}$.

Let's find each derivative:
\begin{align*}
	\frac{dA}{ds} &= 4\alpha_i \\
	\frac{dB}{ds} &= 2\alpha_i s \\
	\frac{dC}{ds} &= \frac{1}{\frac{m_i^2 + \alpha_i s^2}{\mu^2}} \cdot \frac{2\alpha_i s}{\mu^2} = \frac{\mu^2}{m_i^2 + \alpha_i s^2} \cdot \frac{2\alpha_i s}{\mu^2} = \frac{2\alpha_i s}{m_i^2 + \alpha_i s^2}
\end{align*}
Now, substitute these into the second derivative expression:
\begin{align*}
	\frac{d^2f_i}{ds^2} &= (4\alpha_i)(m_i^2 + \alpha_i s^2)\left(\log\left(\frac{m_i^2 + \alpha_i s^2}{\mu^2}\right) - 1\right) \\
	&+ (4\alpha_i s)(2\alpha_i s)\left(\log\left(\frac{m_i^2 + \alpha_i s^2}{\mu^2}\right) - 1\right) \\
	&+ (4\alpha_i s)(m_i^2 + \alpha_i s^2)\left(\frac{2\alpha_i s}{m_i^2 + \alpha_i s^2}\right)
\end{align*}

We need to find $\frac{\partial^2 V^{(1)}(s)}{\partial s^2}\Big|_{s=0}$. This means evaluating each $\frac{d^2f_i}{ds^2}$ at $s=0$ and summing them up, multiplied by the prefactor $\frac{1}{64\pi^2}$.

When $s=0$:
\begin{itemize}
	\item The term $(m_i^2 + \alpha_i s^2)$ becomes $m_i^2$.
	\item The term $\left(\log\left(\frac{m_i^2 + \alpha_i s^2}{\mu^2}\right) - 1\right)$ becomes $\left(\log\left(\frac{m_i^2}{\mu^2}\right) - 1\right)$.
	\item Any term with an explicit factor of $s$ or $s^2$ will become zero.
\end{itemize}
Evaluating $\frac{d^2f_i}{ds^2}$ at $s=0$:
\begin{itemize}
	\item The first term: $4\alpha_i m_i^2 \left(\log\left(\frac{m_i^2}{\mu^2}\right) - 1\right)$
	\item The second term: $0$ (due to $s$ factors)
	\item The third term: $0$ (due to $s$ factors)
\end{itemize}
So, for each term for $f_i(s)$, the mass correction term is:

$$\delta m^2 = \frac{\partial^2 V^{(1)}(s)}{\partial s^2}\Big|_{s=0} =  \frac{1}{64 \pi^2}  4\alpha_i m_i^2 \left(\log\left(\frac{m_i^2}{\mu^2}\right) - 1\right) $$

$$\delta m^2 = \frac{1}{16 \pi^2}  \alpha_i m_i^2 \left(\log\left(\frac{m_i^2}{\mu^2}\right) - 1\right) $$

\end{document}